\documentstyle[12pt,draft]{article}
\begin{document}
\begin{center}
\Large \bf
%%%%%%%%%%%%%%%%%%%%%%%%%%%%%%%%%%%%%%%%%%%%%%%%%%%%%%%%%%%%%%%
Sensitivity of the nuclear equation of state
\\ towards relativistic effects
 \footnote{Supported
by GSI-Darmstadt and BMFT under contract 06T\"U736.} \\
%%%%%%%%%%%%%%%%%%%%%%%%%%%%%%%%%%%%%%%%%%%%%%%%%%%%%%%%%%%%%%
\vspace*{1.0cm}
\normalsize \bf\large
%%%%%%%%%%%%%%%%%%%%%%%%%%%%%%%%%%%%%%%%%%%%%%%%%%%%%%%%%
Rajeev K. Puri, E. Lehmann, Amand Faessler and S.W. Huang\\
\normalsize\it
%%%%%%%%%%%%%%%%%%%%%%%%%%%%%%%%%%%%%%%%%%%%%%%%%%%%%%%%
\vspace*{0.5cm}
Institut f\"ur Theoretische Physik der Universit\"at T\"ubingen,\\
 Auf der Morgenstelle 14, D72076, T\"ubingen, Germany.\\
\vspace*{1.7cm}
%%%%%%%%%%%%%%%%%%%%%%%%%%%%%%%%%%%%%%%%%%%%%%%%%%%%%%%%%
\today
\vspace*{1.5cm}
\bf\normalsize
%%%%%%%%%%%%%%%%%%%%%%%%%%%%%%%%%%%%%%%%%%%%%%%%%%%%%%%%
{\bf Abstract:}\\
\vspace*{0.8cm}
\begin{minipage}{12.5cm}
We analyze
relativistic effects in transverse momentum
using Quantum Molecular Dynamics [QMD] and its covariant extension
Relativistic Quantum Molecular Dynamics [RQMD]. The strength of the
relativistic effects is found to increase with the bombarding
energy and with an averaged impact parameter. The variation in the
intensity of the relativistic effects with variation in the mass
 of the colliding
nuclei is not systematic. Furthermore, the hard EOS is affected
drastically by the relativistic effects whereas the soft EOS is
affected less.
Our analysis shows that up to the bombarding energy of
1 GeV/nucl., the influence  of relativistic effects
is small. Whereas at higher energies,
relativistic effects become
naturally
very important.
\end{minipage}
\end{center}
\newpage
%%%%%%%%%%%%%%%%%%%%%%%%%%%%%%%%%%%%%%%%%%%%%%%%%%%%%
\baselineskip 24 pt
%%%%%%%%%%%%%%%%%%%%%%%%%%%%%%%%%%%%%%%%%%%%%%%%%%%%%
 In last years,
it
has become clear that for analyzing the heavy ion
collisions at intermediate energies, one needs a Lorentz-invariant
theory [1-6]. Therefore, it is important to study the consequences of the
covariant treatment of the dynamics on observables calculated
with different nuclear {\bf
E}quations {\bf O}f {\bf S}tate [EOS].
One of the most promising quantities
which are sensitive to different EOS's is the transverse
momentum [also called "flow" in the literature].
 Moreover, due to the fact that the
transverse momentum is affected drastically by the model parameters,
by the bombarding energy, by the masses of the
colliding nuclei and by the impact parameters
\cite{molitoris1987},
 \cite{aichelin1991}, it is important   to
study the relativistic effects on the
transverse momentum using  different  bombarding
energies, masses of the colliding nuclei and impact parameters.
%%%%%%%%%%%%%%%%%%%%%%%%%%%%%%%%%%%%%%%%%%%%%%%%%%%%

 The differences of observables obtained between a relativistic
and a non-relativistic treatment have different physical origins.
Some of them are
covariant treatment of the dynamics, relativistic forces,
retardation effects or  meson radiations [1-6]. Relativistic
effects which we are going to discuss are the one arise from  full
covariant treatment of the dynamics [RQMD] \cite{lehmann1994}
compared to QMD results. To keep the influence of the relativistic
mean field away from this analysis, we
do not
use the relativistic
mean field. Rather we generalize the normal Skyrme force in such a way that
it
gains well defined Lorentz properties, e.g. is treated as a
Lorentz-scalar.
%%%%%%%%%%%%%%%%%%%%%%%%%%%%%%%%%%%%%%%%%%%%%%%%%%%%%%%%%%%%%

To analyze the transverse momentum,  we use {\bf
Q}uantum {\bf M}olecular {\bf D}ynamics [QMD] and its covariant
extension  {\bf R}elativistic {\bf
Q}uantum {\bf M}olecular {\bf D}ynamics [RQMD].
The RQMD approach was developed by the Frankfurt group \cite{sorge1989}
and is massively used nowadays at ultra-relativistic energies. We use
here a similar method in order to study relativistic effects at intermediate
energies.
The details of QMD and RQMD can be found in
refs.\cite{aichelin1991}
and [2,3], respectively.
%%%%%%%%%%%%%%%%%%%%%%%%%%%%%%%%%%%%%%%%%%%%%%%%%%%%%%%%%

The QMD model is a semi-classical model where important quantum features
like stochastic scattering,
Pauli-principle, creation and reabsorption of the resonances,
 particle production
etc. are also considered. The particles are propagated using
the classical Hamiltonian. During the propagation particles can also
collide elastically or inelastically \cite{aichelin1991}.
The RQMD approach
contains a multi-time description and
propagates the particles in a covariant
fashion. This covariant propagation is coupled with the above
mentioned  quantum features [2,3].    The main notable
differences between QMD and RQMD are:
%%%%%%%%%%%%%%%%%%%%%%%%%%%%%%%%%%%%%%%%%%%%%%%%%%%%%%%%%%%%

\fbox{1}. In RQMD, we have initially a Lorentz-contracted
distribution in coordinate
space and an elongated distribution in  momentum space.

\fbox{2}. The RQMD formalism
contains a multi-time description
which means that each
particle carries its own time coordinate.
 These different time coordinates of baryons can play an
important role in the time evolution of the heavy ion collisions.
 \cite{lehmann1994}.

\fbox{3}. The RQMD approach contains a full
 covariant treatment of the collision part
and the Pauli-blocking. Here in the present study, we include
in RQMD and QMD $\Delta(1232)$ and $N^*$ (1440) resonances
. In the energy region we are interested,
these resonances are the main dominating resonances.

\fbox{4}.  In RQMD, the mean field is a
Lorentz-scalar whereas in QMD it is a zero component of the Lorentz
vector. Due to the covariant feature, the interactions in RQMD are
defined as a function of the distance between the particles in
their centre-of-mass system. Therefore, in a moving frame
 these interactions are  not
 spherical but are Lorentz contracted in the direction of the motion
 of the two interacting particles. Hence,
  the strength of the interaction depends
 strongly on the direction of the center-of-mass motion of the two
 nucleons in the rest frame of the two nuclei. When the initial
coordinate  space
distribution is Lorentz contracted then, naturally, the density of a
 fast moving nucleus becomes much larger than the normal one. If
 one implements this feature
in normal QMD then one finds
that it can lead to a tremendous enhancement in the
transverse flow.
 When one uses this
feature in covariant RQMD,  this artificial repulsion due
to the initial contraction of the coordinate space is partially
counterbalanced and thus
the quantities like density or flow in RQMD lies between  normal
QMD and QMD with a contracted initial coordinate space distribution
 \cite{lehmann1994}.
%%%%%%%%%%%%%%%%%%%%%%%%%%%%%%%%%%%%%%%%%%%%%%%%%%%%%%%

To study the transverse flow, one oftenly takes the average-in-plane
transverse momentum which is defined as:
\begin{equation}
<P_x^{dir}> = \frac{1}{N} \sum_{i=1}^{N}
 \mbox{sign} [Y_i]~~ P_i^x,
\end{equation}
where $Y_i$ and $P_i^x$ are the rapidity and transverse momentum of the
i-th particle, respectively.
%%%%%%%%%%%%%%%%%%%%%%%%%%%%%%%%%%%%%%%%%%%%%%

 The evolution of $<P_x^{dir}>$  flow has been studied  and discussed
 extensivly in past several years [7-8].
 Here our interest is to study the influence
of a covariant formalism compared to a non-covariant
formalism. Therefore, instead of looking to the
absolute values of the flow, we construct another
 quantity which can give us
a kind of measure of
the
{\bf S}ize {\bf O}f {\bf R}elativistic {\bf E}ffects
[SORE] :
\begin{equation}
P_x^{SORE} = \left | \frac{<P_x^{dir}> [RQMD] - <P_x^{dir}> [QMD]}{
                   <P_x^{dir}> [QMD]}\right|.
\end{equation}
The main advantage of $P_x^{SORE}$ is that it
gives us a direct measure of either the enhancement or the reduction in the
flow using RQMD with respect to the QMD. To treat the
enhancement and the
reduction at equal footing, we take the absolute value of $P_x^{SORE}$.
%%%%%%%%%%%%%%%%%%%%%%%%%%%%%%%%%%%%%%%%%%%%%

In case of a covariant formalism, the important factors are the Lorentz
properties which contain the factor $\gamma$ [ = $ \frac{1}{
\sqrt{(1-\beta^2)}}$, where $\beta$ is the velocity of the
particle as compared to the velocity of light]
 and the different time coordinates
of the baryons. Therefore, in fig. 1, we study the $P_x^{SORE}$
( obtained at the final stage of the reaction) as a function of the
bombarding energy E$_{lab}$. It is impressive to note that at 50 MeV/nucl.,
 RQMD and QMD show good agreement.
The  relativistic effects
increase as expected with increasing  bombarding energy. This is true
for both the hard and the soft EOS's. One also sees that up to 1 GeV/nucl.,
the intensity of these relativistic effects is very small. As the flow
is a very sensitive quantity, one has to consider relativistic
effects only if $<P_x^{SORE}>$ lies above  $20 \%$. Therefore, one can use
 non-covariant models up to 1 GeV/nucl. safely. But at higher energies
one sees clearly strong relativistic effects in the transverse momentum.
 We also note that these effects do not
grow in a simple  systematic way as a function  of
 $\gamma^{rel}_{cm}$ $[= \gamma_{cm} -
 \gamma_{cm}(\beta=0)]$ or $\beta_{cm}$.
Thus  the dynamical relativistic effects are not only due
 to compression which is created by the initial
Lorentz contraction of the coordinate space distribution.
%%%%%%%%%%%%%%%%%%%%%%%%%%%%%%%%%%%%%%%%%%%%%%%

To study the  relativistic effects
as a function of the impact parameter, we choose the bombarding energy
of 1.5 GeV/nucl.. At this energy
 one has   medium size relativistic effects ( see fig.1).
In fig. 2 we plot  $P_x^{SORE}$ as a function of the
b$_{scale} [ = b/(R_T+R_P)$, where $R_T$ and $R_P$ are
the radius of the target and projectile, respectively].
 It is interesting to see that  both soft and hard EOS's show
a similar behaviour i.e. the size of relativistic
effects is less for the central collisions. After reaching a
maximum value,  the relativistic effects start
to decrease with increasing impact parameter.
 The main notable results are :
(1) The maximum relativistic effects for the hard EOS are about 170$\%$
whereas for the soft EOS, they are  less than 90$\%$.
 (2) The value of the
$P_x^{SORE}$  as a function of the
impact parameter varies between 26 $\%$ to 170 $\%$.
This
clearly stresses that in order to establish some effects, one has
to take care of the whole impact parameter range.
%%%%%%%%%%%%%%%%%%%%%

To confront the calculations with experimental data,
 one has to average over
a certain impact parameter range.
One often averages by giving equal
weights to all impact parameters (like in the case of the radial flow
analysis of the FOPI collaboration \cite{reisdorf1994}).

 Therefore to establish an average size of
  relativistic effects, we  do a similiar
 averaging. Hence, we define an averaged $<P_x^{SORE}>$ as
\begin{equation}
<P_x^{SORE}> = \left | \frac{\sum_{b_{scale}}<P_x^{dir}>
 [RQMD] - \sum_{b_{scale}}<P_x^{dir}> [QMD]}{
                   \sum_{b_{scale}}<P_x^{dir}> [QMD]}\right|.
\end{equation}
This averaged quantity $<P_x^{SORE}>$ gives us the size of the
relativistic effects  averaged over an
impact parameter range chosen.
We averaged differently : (i) First we averaged  over
so called {\bf C}ollision {\bf D}ominated impact parameters
 i.e. $ 0 < b_{scale} < 0.5$
( defined as  $<P_x^{SORE}>_{CD}$).
 (ii) Second we averaged over the impact parameter range
  $ 0.5 \le b_{scale} \le 1.0$. This
region is dominated by the {\bf M}utual {\bf I}nteractions
 [self-consistent field, the number of collisions is small].
We label this as  $<P_x^{SORE}>_{MID}$.
  From fig. 2, one can see that the
size of  $<P_x^{SORE}>_{CD}$ ( dotted lines)
 is about 49$\%$ for
the hard EOS and about 36 $\%$ for the soft EOS. The mutual interactions
dominated average [$<P_x^{SORE}>_{MID}$] shows very large effect
 (dashed lines) [i.e. 135.8$\%$ effect using hard EOS and 85.6$\%$ using
the soft EOS]. This can be understood by the fact that in case
of  central and semi-central collisions, the self consistent
mean field  does
not play an important role. Whereas for peripheral
collisions, the importance of the covariant treatment of the
dynamics increases. The interesting point is that
 when one averages over
the full impact parameter range [i.e.
 $ 0.0 < b_{scale} \le 1.0$],
one finds that  the hard EOS shows about a 90$\%$
effect whereas the soft EOS shows a less than  50 $\%$. This is remarkable.
The hard and soft EOS's show a very different sensitivity towards
the covariant treatment of the dynamics. This also indicates that
 a full covariant treatment gets larger differences in the
transverse flow using
the hard and the soft EOS's which in a non-covariant model
are always small. This different response of different EOS's towards
relativistic effects stresses that one cannot estimate
the size of  relativistic effects by
 rescaling  the results
obtained in a non-covariant approach.
%%%%%%%%%%%%%%%%%%%%%%%%%%%%%%%%%%%%%%%%%%%%

One of the important features of the Constraint Hamiltonian Dynamics
is that it
contains a multi-time description.
In this formalism, each
particle carries its own time
co-ordinate in a certain frame of reference.
The maximum time difference
in the time coordinates of particles at a fixed bombarding energy
is roughly proportional to the radius of the nucleus.
Thus the ratio of the maximum time difference
to the radii of the nuclei at a given bombarding energy
should be roughly the same for all masses. Therefore, the study
of the relativistic effects using different colliding nuclei
 will give us the possibility to look
whether relativistic effects are due to different time
coordinates only.
If one gets similar effects for all masses then, naturally, the
different time coordinates of particles will be
the only cause for the relativistic effects. Therefore,
in figure 3, we plot the $P_x^{SORE}$ as a function of the masses of
the colliding nuclei. Note that here the bombarding energy (and
hence the $\gamma$)
is the same for all mass systems. Due to fact that RQMD simulations
take extensive computer time, we have to restrict ourselves for masses
A = 80 only. The statistical error in these calculations is about
2\% for light systems (like C-C) and about 4\% for the heaviest
system (Ca-Ca).
One clearly sees that the size of
the relativistic effects
differs appreciably. This  shows that the relativistic effects in
the present study are
not only
created by the multi-time description,
but they
can be due  to a number of
other causes which have been discussed  above.
 Nevertheless the different time
coordinates of the baryons play an important role in heavy ion
collisions.

Concluding, we have analyzed the strength of the relativistic effects
 in the transverse flow by studying  a  variety of reactions using RQMD
and QMD. The relativistic effects which are originating
from the
covariant treatment of the formalism are found to grow with the bombarding\
energy and also with the impact parameter. But we do not see any
monotonic
dependence of the relativistic effects on the masses of the colliding nuclei.
These
results clearly show that the relativistic effects are not only due
to the compression
by Lorentz contaction
or due to the different time coordinates of the
particles but, there are a number of physical causes which can produce these
effects. An other important conclusion is that the hard and the soft EOS's show
different sensitivities towards relativistic effects. One finds
in a  covariant calculation far more differences
 in the transverse flow using the
 the hard and the soft EOS's than in the non-covariant QMD.
 %%%%%%%%%%%%%%%%%%%%%%%%%%%%%%%%%%%
\vspace*{2mm}\\

 The authors are thankful to Prof. J\"org Aichelin for providing
his latest QMD code.

 \newpage
{\large\bf Figure Captions}\\
\vspace*{0.8cm}\\
{\bf Fig.1} The size of the relativistic effects
$P_x^{SORE} \%$ (eq. 2) as a function of the bombarding energy. The
reaction under consideration is $^{40}$Ca-$^{40}$Ca at an impact
parameter  2 fm. The variation of the $\gamma_{cm}^{rel} \%$
[i.e. $\gamma_{cm}^{rel}$ x 100] and
$\beta_{cm} \%$ [$\beta_{cm} $ x100 ]
as a function of the bombarding energy is also shown. Here
$\gamma_{cm}^{rel}$ is defined as $\gamma_{cm} -1$.  \\
\vspace*{0.8cm} \\
{\bf Fig.2} The $P_x^{SORE} \%$ as a function of the scaled impact parameter
$b_{scale}$. For the definition of the
$<P_x^{SORE}>_{CD}$, $<P_x^{SORE}>_{MID}$,
 see  text.\\
 \vspace*{0.8cm}\\
 {\bf Fig.3}. The same as in fig. 1 but $P_x^{SORE} \%$ is
plotted as a
function of mass of the colliding nuclei. Here the impact parameter
is b$_{scale}$ = 0.25 b$_{max}$ [$b_{max}$ = $ R_T + R_P$]
 and the bombarding energy 1.5 GeV/nucl..
\end{document}